\documentclass{article}
\usepackage{spconf,amsmath,graphicx}

\usepackage{enumitem}
\setlist{nosep, leftmargin=14pt}

\usepackage{amssymb} 
\usepackage{booktabs}
\usepackage{multirow}

\usepackage[belowskip=-6pt,aboveskip=0pt]{caption}

\DeclareRobustCommand{\eg}{\emph{e.g. }}

\DeclareRobustCommand{\etal}{{et al. }}

\title{{TranSOP: Transformer-based Multimodal Classification for Stroke Treatment Outcome Prediction}}
\name{Author(s) Name(s)\thanks{Some author footnote.}}
\address{Author Affiliation(s)}

\name{Zeynel A. Samak$^{1}$ \qquad Philip Clatworthy$^{2,3}$ \qquad Majid Mirmehdi$^{1}$}

\address{$^{1}$ Department of Computer Science, University of Bristol, Bristol, UK \\ $^{2}$ Translational Health Sciences, University of Bristol, Bristol, UK
\\ $^{3}$  Stroke Neurology, Southmead Hospital, North Bristol NHS Trust, Bristol, UK}

\begin{document}
%
\maketitle

\begin{abstract}
Acute ischaemic stroke, caused by an interruption in blood flow to brain tissue, is a leading cause of disability and mortality worldwide. The selection of patients for the most optimal ischaemic stroke treatment is a crucial step for a successful outcome, as the effect of treatment highly depends on the time to treatment.
We propose a transformer-based multimodal network (TranSOP) for a classification approach that employs clinical metadata and imaging information, acquired on hospital admission, to predict the functional outcome of stroke treatment based on the modified Rankin Scale {(mRS)}. This includes a fusion module to efficiently combine 3D {non-contrast computed tomography (NCCT)} features and clinical information. In comparative experiments using unimodal and multimodal data on the MRCLEAN dataset, we achieve a state-of-the-art AUC score of 0.85.
\end{abstract}

\section{Introduction}
\label{sec:intro}

Acute ischaemic stroke is the most common type of stroke and a leading cause of disability and mortality worldwide \cite{neethi2022stroke}. It is a condition caused by the formation of clots, following interruption of blood flow to the brain. If the blockage is not resolved, the extent of dead tissue increases and the irreversible ischaemic core expands over time. As Saver \cite{saver2006time} stated, "Time is brain" for stroke diagnosis and treatment, and it is essential to carry out the appropriate treatment in a timely manner. Although thrombectomy is the most effective treatment for ischaemic stroke cases, there is a risk of brain haemorrhage and death. Therefore, determining if a patient just admitted can benefit from mechanical thrombectomy leading to a good functional outcome, is an important step towards reducing risk and improving the quality of life for stroke patients.

Methods for automatic outcome prediction of stroke treatment have been proposed using logistic regression \cite{venema2017selection,ramos2020predicting}, random forests \cite{vanOs2018,Nawabi2021}, support vector machines \cite{ramos2020predicting,hofmeister2020clot}, and recently, convolutional neural networks (CNNs) \cite{hilbert2019data,Bacchi2019DeepStudy,SAMAK2022102089}. Some use clinical records \cite{venema2017selection,vanOs2018,ramos2020predicting}, imaging information \cite{hilbert2019data,hofmeister2020clot,Nawabi2021}, or a combination of both \cite{Bacchi2019DeepStudy,samak2020prediction,kappelhof2021evolutionary}. The CNN-based models have been applied to various imaging modalities, \eg magnetic resonance imaging (MRI), NCCT and CT angiography (CTA). While such deep learning models perform well in medical image analysis, 3D CNN models that exploit 3D brain volumes require numerous parameters and computational resources. Furthermore, they cannot learn long-range relationships due to their limited receptive field. In contrast, more recently, transformers have achieved outstanding results in various applications thanks to their big data and model size scalability and better longer-range attention-based modelling capability \cite{dosovitskiy2020image,khan2021transformers}.  However, pure transformer-based methods have not been widely applied in medical image classification due to their limited performance on small datasets \cite{jang2022m3t}. 

\begin{figure}[t]
	\centering
\includegraphics[width=\linewidth]{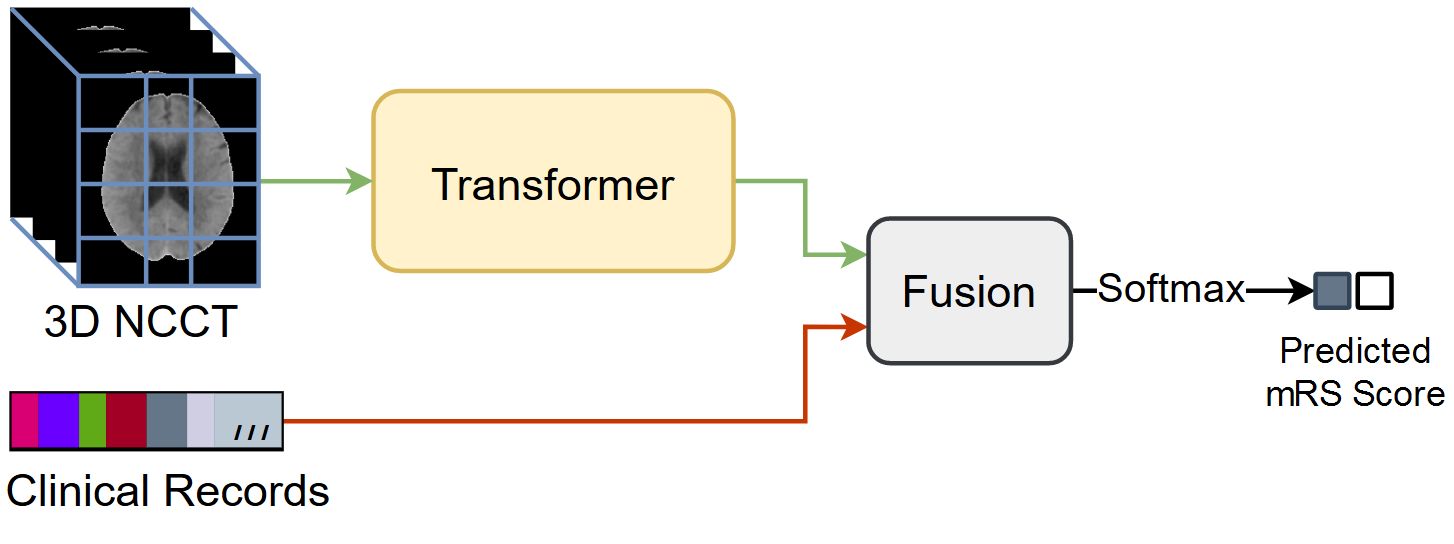}
	\caption{TranSOP predicts functional outcome of ischaemic stroke treatment leveraging only the baseline NCCT scan and clinical records available on hospital admission.}
	\label{FIG:intro_fig}
\end{figure}

In this paper, we introduce TranSOP, a transformer-based multimodal architecture to predict functional outcomes of ischaemic stroke patients 90 days after treatment (see Fig  \ref{FIG:intro_fig}). We combine clinical metadata (\eg gender, age, hypertension, glucose level) and 3D NCCT obtained at the point of hospital admission for 500 ischaemic stroke patients. We also explore different strategies for this multimodal fusion and conduct extensive experiments on various architectures, including ViT, ViT with CNN, pre-trained ViT (from DeiT \cite{touvron2021training}) and Swin transformer (SwinT) \cite{liu2021swin} in our TranSOP model. 

\begin{figure*}[t]
    \centering
    \includegraphics[width=\textwidth]{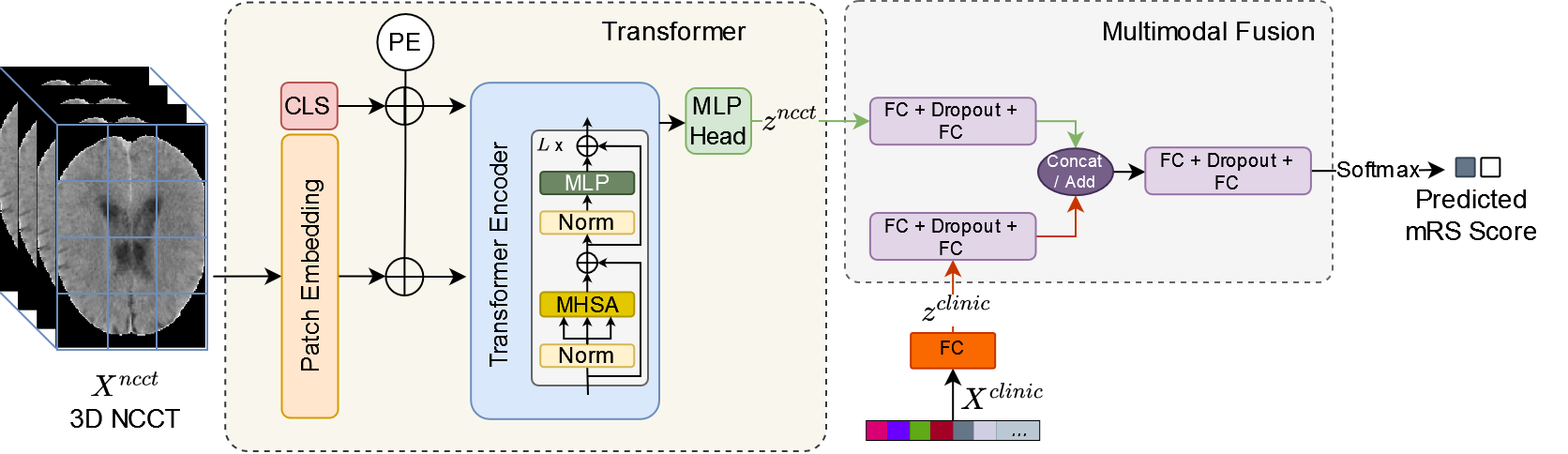}
    \caption{Overview of our proposed transformer-based multimodal architecture, TranSOP. PE: positional encoding, CLS: a token/vector that represents the input volume for classification, MHSA: Multi-head self-attention, MLP: multi-layer perceptron, FC: fully connected layer.}
    \label{figc6:trans_modal}
\end{figure*}

\section{Related Works}
\label{sec:related}

{There are only a few studies that have employed CNN-based multimodal networks to predict the functional outcome of stroke treatments, \eg for thrombolysis \cite{Bacchi2019DeepStudy} and for thrombectomy \cite{samak2020prediction,SAMAK2022102089}. Bacchi \etal \cite{Bacchi2019DeepStudy}  applied a CNN model to 3D NCCT images and clinical records of patients who underwent thrombolysis treatment. {Samak \etal \cite{samak2020prediction} also proposed a multimodal CNN architecture with channel-wise and spatial attentional blocks to predict dichotomised mRS scores from baseline 3D NCCT scans and clinical records of MR CLEAN \cite{fransen2014mr} dataset}. {Further, in  \cite{SAMAK2022102089}, Samak \etal }additionally incorporate 1-week follow-up scans during their model training to encode stroke changes over time for better mRS score prediction.} 

Transformers have shown significant success in natural language processing, \eg machine translation \cite{vaswani2017attention}, and computer vision, \eg medical imaging tasks \cite{dai2021transmed,hatamizadeh2022swin}. They facilitate a mechanism of self-attention that can model the long-range dependency of sequences and focus on important features. Dosovitskiy \etal \cite{dosovitskiy2020image} proposed the first pure vision transformer (ViT), applied directly to sequences of image patches for image classification. ViTs have obtained comparable and even better results in some tasks than CNNs, \eg for object detection \cite{carion2020end,wang2022fpdetr}.

Since its introduction ViT has been deployed in medical image segmentation using different imaging modalities. UNETR \cite{hatamizadeh2022unetr} adapts the commonly deployed and successful U-Net architecture \cite{cciccek20163d}, by replacing its convolutional encoder with a transformer encoder and modifying its convolutional decoder based on the output of the transformer encoder for image segmentation. Similarly, other studies  \cite{hatamizadeh2022unetformer,hatamizadeh2022swin,li2022transiam} also replace the convolutional encoder with a transformer encoder, while some integrate the transformer encoder into the bottleneck of a U-Net-like model \cite{luo2021ucatr,wang2021transbts,ji2021multi,wang2022metrans} or use hybrid blocks that combine the convolutional and transformer layers \cite{gao2021utnet,liu2022transfusion}. Such works have been applied to NCCT \cite{luo2021ucatr}, MRI \cite{wang2021transbts,gao2021utnet,hatamizadeh2022swin,liu2022transfusion} and microscope \cite{ji2021multi} images. In another recent work, Amador \etal \cite{amador2022hybrid} propose a hybrid model that performs segmentation of the final lesion outcome {of ischaemic stroke} from {baseline} spatio-temporal CT perfusion (CTP) images using a transformer encoder embedded in the U-Net bottleneck. 

Although most transformer-based models in medical image analysis are in the {\it segmentation} domain, there are some studies that have employed them on medical image {\it classification}, \eg for COVID-19 \cite{mondal2021xvitcos,park2022multi}, retinal disease \cite{wu2021vision}, cell analysis \cite{wang2021transpath,gao2021instance}, brain tumour \cite{dai2021transmed,jun2021medical}, Alzheimer's disease \cite{li2022trans,jang2022m3t} classification and age estimation \cite{jun2021medical,he2021global}. These methods are based on a pure transformer \cite{mondal2021xvitcos,yu2021mil,matsoukas2021time,gheflati2022vision,bhattacharya2022radiotransformer} or a hybrid model that uses ResNet \cite{dai2021transmed,li2022trans,wu2021vision}, DenseNet \cite{park2022multi} or a CNN module \cite{wang2021transpath,sriram2021covid,gao2021instance,he2021global,jang2022m3t} followed by a transformer encoder on 2D imaging modalities like X-Rays \cite{bhattacharya2022radiotransformer,jacenkow2022indication}, microscope images  \cite{wang2021transpath,gao2021instance} and 3D MRI volumes \cite{dai2021transmed,li2022trans,jang2022m3t,jun2021medical}. To the best of our knowledge, there are no studies using the transformer in 3D NCCT classification and prediction of functional stroke outcomes from unimodal or multimodal data.

\section{Proposed Method}
An overview of the proposed architecture, TranSOP, is shown in Fig. \ref{figc6:trans_modal}, which includes a transformer encoder and a multimodal fusion module to predict mRS scores. 
Transformers can process 1D input sequences, as originally used in the NLP domain where each word is embedded in a 1D vector as a token. Similarly, we split a 3D NCCT volume, $X^{ncct}_i \in \mathbb{R}^{1 \times D \times W \times H}$, into 1D vectors via patch embedding where $D$, $W$, and $H$ are depth, width and height, and a volume is divided into non-overlapping patches of size $P^3$, which generate a sequence of 1D patch vectors of length $L=[\frac{D}{P}] \times [\frac{W}{P}] \times [\frac{H}{P}]$.

We use a convolutional layer to project each patch into a $K$ dimensional embedding space \cite{touvron2021training,hatamizadeh2022swin}. We add a learnable parameter $[CLS] \in \mathbb{R}^{1 \times K}$, to the patch embedding sequence to represent the entire volume for classification. In addition, a learnable positional encoding, ($PE \in \mathbb{R}^{(L+1) \times K}$) is added to the sequences, so that the spatial information of the patches can be preserved (see Fig. \ref{figc6:trans_modal}). Next, a series of transformer blocks, each including a normalisation layer followed by multi-head self-attention (MHSA), a normalisation layer, and a multi-layer perceptron (MLP) head are utilised in the {transformer encoder}.  Then, an MLP head is applied to the classification token to extract NCCT volume features $z^{ncct}$ for the fusion process. Clinical metadata features $z^{clinic}$ are computed by a fully connected layer (FC) (orange box in Fig. \ref{figc6:trans_modal}). 
 
{In the multimodal fusion module, a stack of two FCs with a dropout layer in-between prepare the input scan's $z^{ncct}$ and $z^{clinic}$ for fusion  (see right box in Fig. \ref{figc6:trans_modal}).} We use two methods for the fusion of these image volume and clinical features, (i) concatenation where both features are joined to make a larger 1D vector and (ii) addition where both features are added element-wise with each feature vector  multiplied by a learnable weight. Finally, another stack of FC, Dropout, and FC layers is applied to the fused features before being passed to a  \textit{Softmax} layer for final predictions. These predictions are dichotomised mRS scores, where mRS $\leq$ 2 indicates a good outcome and mRS $>$ 2 expresses a bad outcome. Note that, dropout layers are deactivated during inference.

\section{Experiments \& Results}
\label{sec:exp}

{\bfseries Dataset --} We used the MR CLEAN Trial dataset\footnote{https://www.mrclean-trial.org/home.html}, collected from a multi-centre study, which is one of the most comprehensive datasets of patients who underwent ischaemic stroke treatment. 
Five hundred patients (233 assigned to mechanical thrombectomy and 267 to usual care) were treated in 16 medical centres in the Netherlands. We refer the reader to the MR CLEAN study protocol \cite{Berkhemer2015AStroke,fransen2014mr} for more detailed information on the dataset.

Through pre-processing, some of the apparent variations due to various acquisition protocols at different clinical centres were reduced to allow our model to deal with more similar standard input. First, all scans were re-sampled to the same voxel size of 3x1x1$mm^3$, followed by clipping the intensity range of 0-80HU. The skull structure was then removed in the NCCT scans and the volumes were cropped to $32\times192\times128$ from the centre.

Data augmentations, such as horizontal/vertical flips and Gaussian noise, were applied to increase the variation and number of input samples to help improve the robustness of the network.  Finally, the voxels of the NCCT scans were normalised to zero mean and one standard deviation.

{\bfseries Implementation Details --} We split the dataset into three subsets, training (70\%, 350 patients), validation (15\%, 75 patients) and testing (15\%, 75 patients). The proposed model was trained for 500 epochs using an Adam optimiser with a weight decay of 0.0001, a learning rate of 0.0003, and a batch size of 24. A cosine learning rate scheduler was used. The experiments were implemented in PyTorch and MONAI \cite{monai2020} on a single NVIDIA P100 16GB GPU.

{\bfseries Details of Experiments --} We evaluated the performance of our proposed approach against  two existing methods and various transformer architectures that also operate on 3D NCCT volumes and predict the functional outcome of stroke treatment.   The methods of Bacchi \etal \cite{Bacchi2019DeepStudy} and {Samak \etal \cite{samak2020prediction}} {which both use imaging and clinical information,} were re-trained on the registered MR CLEAN dataset and our data split from scratch. {Although, the {FeMA \cite{SAMAK2022102089}} model performs a similar task, it additionally uses 1-week follow-up scans that contain information on stroke changes after treatment during model training. Hence, in the interest of direct comparability, we do not include that work in the present evaluation.}

{We also evaluated our TranSOP approach using different transformer architectures for its encoder part. These are referred to as TranSOP$_{ViT}$, TranSOP$_{DeiT}$, TranSOP$_{ConViT}$ and TranSOP$_{SwinT}$.  TranSOP$_{ViT}$ uses the ViT network and is trained from scratch, TranSOP$_{DeiT}$ utilises the ImageNet pre-trained DeiT model to demonstrate the effect of transfer learning, and TranSOP$_{ConViT}$ uses the first three layers of convolutional blocks before the input is fed into the ViT model to explore the performance of a hybrid model. These three models have the same ViT network which consist of 12 layers of transformer blocks, 12 heads, a hidden MLP feature size of 768 and 3072. In TranSOP$_{SwinT}$, four stages each consisting of two Swin transformer blocks and $N$ MHSA heads, where $N=\{3, 6, 12, 24\}$ for each stage respectively, were used. The {ClinicDNN} model only consumed clinical information to show the expected benefit from imaging information.} {Note, the multimodal fusion step is the same for all the models.}

\begin{table*}
\caption{Results of the models with and without clinical records. The best and second best results are shown in bold and underlined respectively. The second and third rows are convolutional-based models.  CI is confidence interval.} 
\label{tablec6:init_result_end2end2}
\centering

\scalebox{0.91}{
\begingroup
\renewcommand{\arraystretch}{1.} 
\begin{tabular}{lcccclccc}
\toprule
& \multicolumn{3}{c}{\textbf{w/o Clinical Records}} & \phantom{a}& \multirow{2}{*}{\textbf{Fusion}} & \multicolumn{3}{c}{\textbf{with Clinical Records}} \\ 
\cmidrule{2-4} \cmidrule{7-9}
 \textbf{Method}  & \textbf{ACC {\scriptsize (95\% CI)}} & \textbf{F1-score {\scriptsize (95\% CI)}} & \textbf{AUC {\scriptsize (95\% CI)}} &  & & \textbf{ACC {\scriptsize (95\% CI)}} & \textbf{F1-score {\scriptsize (95\% CI)}} & \textbf{AUC {\scriptsize (95\% CI)}}  \\
 \midrule
  ClinicDNN\textsuperscript{\textasteriskcentered} &-&-&-&& -& 0.75 {\scriptsize (0.65-0.85)} & 0.44 {\scriptsize (0.19-0.64)} & 0.73 {\scriptsize (0.57-0.86)} \\ 
  \midrule
 \multirow{2}{*}{{Samak \etal \cite{samak2020prediction}}} & 
          \multirow{2}{*}{\underline{0.72} {\scriptsize (0.62-0.82)}} & 
          \multirow{2}{*}{{0.33} {\scriptsize (0.09-0.53)}} & 
          \multirow{2}{*}{{0.63} {\scriptsize (0.44-0.81) }} & 
          \multirow{2}{*}{} & concat&
          \multirow{1}{*}{{0.77} {\scriptsize (0.66-0.87)}} & 
          \multirow{1}{*}{{0.47} {\scriptsize (0.18-0.67)}} &
          \multirow{1}{*}{{0.78} {\scriptsize (0.63-0.91)}} \\
          & &  & &&add& \underline{0.79} {\scriptsize (0.69-0.89)} & {0.44} {\scriptsize (0.17-0.67)} & {0.71} {\scriptsize (0.51-0.88)}  \\
\midrule

 \multirow{2}{*}{{Bacchi \etal \cite{Bacchi2019DeepStudy}}} & 
          \multirow{2}{*}{\textbf{0.75} {\scriptsize (0.65-0.85)}} & 
          \multirow{2}{*}{{0.40} {\scriptsize (0.16-0.60)}} & 
          \multirow{2}{*}{\underline{0.66} {\scriptsize (0.48-0.80)}} & 
          \multirow{2}{*}{} & concat&
          \multirow{1}{*}{{0.73} {\scriptsize (0.62-0.83)}} & 
          \multirow{1}{*}{{0.51} {\scriptsize (0.29-0.68)}} &
          \multirow{1}{*}{{0.78} {\scriptsize (0.62-0.90)}} \\
          & &  & &&add& {0.73} {\scriptsize (0.62-0.83)} & {0.51} {\scriptsize (0.29-0.68)} & {0.78} {\scriptsize (0.62-0.90)}  \\

\midrule
 \multirow{2}{*}{{TranSOP$_{ConViT}$}} & 
          \multirow{2}{*}{{0.58} {\scriptsize (0.46-0.69)}} & 
          \multirow{2}{*}{{0.40} {\scriptsize (0.21-0.56)}} & 
          \multirow{2}{*}{\textbf{0.67} {\scriptsize (0.46-0.85)}} & 
          \multirow{2}{*}{} & concat&
          \multirow{1}{*}{{0.77} {\scriptsize (0.68-0.87)}} & 
          \multirow{1}{*}{\underline{0.58} {\scriptsize (0.36-0.74)}} &
          \multirow{1}{*}{{0.83} {\scriptsize (0.72-0.93)}} \\
          & &  & &&add& {0.77} {\scriptsize (0.68-0.87)} & \underline{0.58} {\scriptsize (0.36-0.74)} & {0.82} {\scriptsize (0.71-0.92)}  \\
\midrule

 \multirow{2}{*}{{TranSOP$_{DeiT}$}} & 
          \multirow{2}{*}{{0.58} {\scriptsize (0.46-0.69)}} & 
          \multirow{2}{*}{{0.40} {\scriptsize (0.21-0.56)}} & 
          \multirow{2}{*}{{0.63} {\scriptsize (0.44-0.80)}} & 
          \multirow{2}{*}{} & concat&
          \multirow{1}{*}{{0.77} {\scriptsize (0.68-0.86)}} & 
          \multirow{1}{*}{{0.53} {\scriptsize 0.30-0.71)}} &
          \multirow{1}{*}{{0.82} {\scriptsize (0.68-0.93)}} \\
          & &  & &&add& \underline{0.79} {\scriptsize (0.69-0.89)} & {0.52} {\scriptsize (0.27-0.71)} & \underline{0.84} {\scriptsize (0.71-0.94)}  \\
\midrule
 \multirow{2}{*}{{TranSOP$_{ViT}$}} & 
          \multirow{2}{*}{{0.58} {\scriptsize (0.46-0.69)}} & 
          \multirow{2}{*}{{0.40} {\scriptsize (0.21-0.56)}} & 
          \multirow{2}{*}{{0.60} {\scriptsize (0.40-0.78)}} & 
          \multirow{2}{*}{} & concat&
          \multirow{1}{*}{\textbf{0.80} {\scriptsize (0.70-0.89)}} & 
          \multirow{1}{*}{{0.53} {\scriptsize (0.28-0.74)}} &
          \multirow{1}{*}{\underline{0.84} {\scriptsize (0.72-0.94)}} \\
          & &  & &&add& \textbf{0.80} {\scriptsize (0.70-0.89)} & \textbf{0.59} {\scriptsize (0.35-0.76)} & {0.83} {\scriptsize (0.71-0.93)}  \\
\midrule
 \multirow{2}{*}{{TranSOP$_{SwinT}$}} & 
          \multirow{2}{*}{{0.58} {\scriptsize (0.46-0.69)}} & 
          \multirow{2}{*}{{0.40} {\scriptsize (0.21-0.56)}} & 
          \multirow{2}{*}{{0.64} {\scriptsize (0.44-0.82) }} & 
          \multirow{2}{*}{} & concat&
          \multirow{1}{*}{{0.76} {\scriptsize (0.66-0.86)}} & 
          \multirow{1}{*}{{0.54} {\scriptsize (0.32-0.71)}} &
          \multirow{1}{*}{{0.83} {\scriptsize (0.71-0.93)}} \\
          & &  & &&add& \underline{0.79} {\scriptsize (0.69-0.89)} & {0.55} {\scriptsize (0.31-0.73)}& \textbf{0.85} {\scriptsize (0.75-0.94)} \\

\bottomrule
\multicolumn{9}{l}{{\textsuperscript{\textasteriskcentered} A method that uses only clinical metadata information.}} \\
\end{tabular}
\endgroup
}
\end{table*}

We evaluated the classification performance of the models with three commonly used metrics, Accuracy, {\it F1-score} and Area Under ROC Curve (AUC). 
{Table \ref{tablec6:init_result_end2end2} reports the evaluations of the transformer-based and convolution-based networks, along with confidence intervals, for two fusion methods. Broadly, the CNN-based state of the art works \cite{Bacchi2019DeepStudy,samak2020prediction} outperformed the transformer methods when only imaging information was used, for example, \cite{Bacchi2019DeepStudy} and \cite{samak2020prediction} performed best and second best in accuracy at 0.75 and 0.72 respectively. On the other hand, transformer-based methods exceeded Bacchi \etal \cite{Bacchi2019DeepStudy} and Samak \etal \cite{samak2020prediction} when clinical records were included for multimodal analysis, with the best result obtained by TranSOP$_{SwinT}$ at 0.85 AUC. These variations in performance by the transformer could be attributed to both the transformer's appetite for  larger datasets (see \cite{khan2021transformers}), and its already established superiority in handling 1D natural language data.} {As TranSOP$_{SwinT}$ achieved the best AUC score, and it is more efficient thanks to its hierarchical architecture and shifted windowing, it can be a more preferable approach.}

{The results on the use of fusion methods (concat and addition) are inconclusive and further investigation on more efficient fusion methods is necessary.}  

\section{Conclusion}
\label{sec:conclusion}
In this work, we investigated the performance of various networks in predicting the functional outcome of ischaemic stroke treatment based on 3D NCCT scans and clinical information, such as age, sex, and demographic data from the patient's medical history records. Transformer models outperformed convolutional architectures in multimodal settings. This suggests that transformer models, although not performing as well on only imaging data, can learn better complementary imaging information when combined with clinical metadata. In future work, we plan to investigate and explore a data-efficient transformer model for small image datasets. In addition, we would like to extend the proposed architecture to use follow-up scans, such as used in the FeMA \cite{SAMAK2022102089} method during model training.

\section{Acknowledgments}
\label{sec:acknowledgments}
The authors would like to thank the Principal Investigators of the MR CLEAN trial: Profs Aad van der Lugt, Diederik W.J. Dippel, Charles B.L.M. Majoie, Yvo B. W.E.M. Roos, Wim H. van Zwam and Robert J. van Oostenbrugge for providing the data.
Z.A. Samak is funded by the Ministry of Education (1416/YLSY), the Republic of Türkiye.

{\small 
\bibliographystyle{IEEEbib}
\bibliography{Template_ISBI_latex}
}
\end{document}